\documentclass[final]{elsart}

\newcommand\degrees[1]{\ensuremath{#1^\circ}}

\usepackage[dvips,final]{graphicx}
\usepackage{palatino}
\usepackage{url}
\usepackage{amssymb}
\usepackage{multirow}

\journal{Journal of the European Ceramic Society}
\begin{document}

  \begin{frontmatter}
    \title{\textbf{\Huge{Prediction of the functional properties of
	  ceramic materials from composition using artificial neural
	  networks}}} %\\ \small{Draft of \today}}

    \author[ccs]{D. J. Scott},
    \author[ccs]{P. V. Coveney\corauthref{cor}},
    \author[ic]{J. A. Kilner},
    \author[ic]{J. C. H. Rossiny},
    \author[lsbu]{N. Mc N. Alford}

    \corauth[cor]{Corresponding author}

    \address[ccs]{Centre for Computational Science,
      Department of Chemistry,
      University College London,
      Christopher Ingold Laboratories,
      20 Gordon Street,
      London,
      WC1H~0AJ}
    \address[ic]{
      Department of Materials,
      Imperial College London,
      Exhibition Road,
      London,
      SW7~2AZ}
    \address[lsbu]{
      Centre for Physical Electronics and Materials,
      Faculty of Engineering, Science and the Built Environment,
      London South Bank University,
      103 Borough Road,
      London,
      SE1~0AA}

    \begin{keyword}
      C. Dielectric properties \sep C. ionic conductivity \sep
      D. perovskites \sep E. functional applications \sep neural
      networks
    \end{keyword}

    \begin{abstract}
      We describe the development of artificial neural networks (ANN)
      for the prediction of the properties of ceramic materials. The
      ceramics studied here include polycrystalline, inorganic,
      non-metallic materials and are investigated on the basis of
      their dielectric and ionic properties. Dielectric materials are
      of interest in telecommunication applications where they are
      used in tuning and filtering equipment. Ionic and mixed
      conductors are the subjects of a concerted effort in the search
      for new materials that can be incorporated into efficient, clean
      electrochemical devices of interest in energy production and
      greenhouse gas reduction applications. Multi-layer perceptron
      ANNs are trained using the back-propagation algorithm and
      utilise data obtained from the literature to learn
      composition-property relationships between the inputs and
      outputs of the system. The trained networks use compositional
      information to predict the relative permittivity and oxygen
      diffusion properties of ceramic materials. The results show that
      ANNs are able to produce accurate predictions of the properties
      of these ceramic materials which can be used to develop
      materials suitable for use in telecommunication and energy
      production applications.
    \end{abstract}
  \end{frontmatter}

  \section{Introduction}
  Accurate determination of the properties of a ceramic
  \cite{bib:IntroductionToCeramics} material allows it to be matched
  to appropriate applications and therefore, fast and reliable methods
  for predicting material properties would be extremely
  useful. Various modelling techniques can be used to predict material
  properties from compositional and processing information and can
  provide fast and accurate results. In the conventional ``Popperian''
  scientific method \cite{bib:ConjectandRefutate}, a theory is
  proposed and tested by experiment. Whilst such experiments increase
  our confidence in the model, one experiment can falsify the theory
  which can never have more than a provisional status. Inductive
  ``Baconian'' methods \cite{bib:BaconNovumOrganum}, in contrast to
  the Popperian technique, begin with experiment and use statistical
  inference to develop a model. In principle, such techniques make no
  prior assumptions about an underlying theoretical model and utilise
  statistical methods to induce data relationships.

  Popperian modelling of the properties of ceramic materials has
  yielded important results. Models of the diffusion of oxygen through
  mixed ionic conductors
  \cite{bib:ModellingAndDatabaseSearchMOICsByCombinatorialMethods}
  have provided accurate predictions of the diffusion coefficient;
  likewise, successful modelling of the structure-performance
  relationship of solid oxide fuel cell (SOFC) electrodes has been
  performed \cite{bib:StructurePropertyRelationshipSOFCCathodes}.
  Prediction of the grain boundary properties of dielectric ceramic
  materials has also been performed
  \cite{bib:ModellingOfGrainBoundaryInBaTiO}. In this paper, by
  contrast, we attempt to produce a Baconian model capable of
  predicting properties of a wide range of ceramic materials. It would
  be extremely difficult to develop such a model using conventional
  (Popperian) techniques \cite{bib:NeuralNetworksInChemistry} and
  therefore, we use the inductive approach known as an artificial
  neural network (ANN).

  ANNs are one of several ``biologically inspired'' computational
  methods which can be used to capture complex, non-linear
  relationships between data
  \cite{bib:NeuralNetworksForPatternRecognition}. These techniques
  have been used in many areas of chemistry
  \cite{bib:NeuralNetworksInChemistry} and have provided accurate
  predictions in the performance of oil-field cements
  \cite{bib:UsingANNstoPredictCement} and materials with highly
  complex behaviour \cite{bib:ModellingHydratingCements}. It is
  generally accepted that ANNs provide more accurate predictive
  capabilities than methods based on traditional linear or non-linear
  statistical regression \cite{bib:PracticalNNRecipesInC++} and the
  superiority of ANNs over regression techniques increases as the
  dimensionality and/or non-linearity of the problem increases
  \cite{bib:ANNsFundamentalsComputingDesignApplication}. ANNs have
  been found to outperform regression techniques in the prediction of
  ceramic material properties
  \cite{bib:InvestigationOfBTANNMethod,bib:PredictionsOfPerovskiteUnitCellLatticeConstants,bib:PredictionsOfPerovskiteLatticeConstants,bib:ModelingAndAnalysisOfPZTByNNs}. Additionally,
  the prediction of dielectric properties of organic materials has
  been attempted \cite{bib:DielectricConstantOrganicLiquidsCompMLR}
  and, again, ANNs have been found to be superior.

  In previously published work, compositional information has formed
  the core of the ANN input data although other descriptors can be
  added to the model to help improve performance
  \cite{bib:InterpretingANNModelsDescriptorImportance,bib:DevelopmentOfSPRForPredictionOfDCUsingNN}.
  Whilst accurate prediction of ceramic material properties has been
  performed previously, this is often based on one material, to which
  dopants are applied. Here we have attempted property prediction of a
  much wider range of materials than completed previously. Materials
  in our datasets range from single compounds, through to complex
  binary, ternary and quaternary systems. The large range of materials
  studied requires statistical techniques capable of handling
  high-dimensional problems; ANNs are ideally suited for this
  purpose. With an automated combinatorial robotic instrument such as
  the London University Search Instrument (LUSI)
  \cite{bib:InstrumentControlAndInformaticsSystem} we can use ANNs to
  rapidly scan compositional parameter space, searching for desirable
  materials.

  \section{Electroceramic materials}
  The study of ceramic materials is a wide ranging and complex subject
  due to both the large range of materials available and the varied
  properties exhibited \cite{bib:Electroceramics}. In our own work, we
  are interested in dielectric ceramics for use in communications
  equipment and oxygen diffusion properties of ceramics for fuel cells
  components. The continuing growth of mobile telecommunications has
  sustained the interest in novel ceramics for use as dielectric
  resonators (DRs) at microwave frequencies (1-20 GHz). New materials
  are constantly required for use in resonators and
  filters. Additionally, ion-diffusing ceramics are employed in a wide
  range of applications. In particular, electrochemical devices such
  as oxygen separation membranes, solid oxide fuel cell (SOFC)
  cathodes, and syngas reactors make use of the ion-diffusion
  properties of ceramic materials.

  One of the most promising classes of materials suitable for use in
  such applications are the perovskite oxides with the general formula
  ABO$_{3}$. A and B are rare earth/alkaline earth ions and transition
  metal cations, respectively. By doping both the A- and B-sites with
  similar metallic elements, the composition of these materials can be
  broadened to encompass a very large number of possible
  combinations. Dopant species and compositions can have a major
  effect on the properties of the material.

  Due to the difficult and time consuming process of conventional
  compound synthesis, scientists are increasingly turning to
  high-throughput combinatorial techniques to develop suitable
  materials \cite{bib:CombSearches,bib:CombApproach}.  Combinatorial
  projects can generate vast quantities of data which require
  informatics and database systems
  \cite{bib:InstrumentControlAndInformaticsSystem} for data entry,
  organisation and data mining. Our Functional Oxide Discovery (FOXD)
  project, which is based around LUSI, aims to utilise artificial
  neural network data analysis techniques to complete the ``materials
  discovery cycle'', allowing the predictions made to direct the
  search for new materials into as yet unexplored territory
  \cite{bib:InstrumentControlAndInformaticsSystem,bib:foxd}.

  \subsection{Microwave dielectric materials for communications equipment}
  The ideal properties of a dielectric resonator (DR) are a
  sufficiently high relative permittivity to allow miniaturisation of
  the component ($\epsilon_{r} > 10$) and high `Q' factor at microwave
  frequencies to improve selectivity (Q $>$ 5000). The quality factor,
  Q is given by the inverse of the dissipation factor $Q= 1/\tan
  \delta$ where $\delta$ is the \emph{loss angle}, the phase shift
  between the voltage and current when an AC field is applied to a
  dielectric material.
  \cite{bib:MicrowaveCeramicsForResonatorsAndFilters}.

  Many useful dielectric resonator materials are perovskites (e.g.
  (Ba,Sr)TiO$_{3}$, (Ba,Mg)TaO$_{3}$, 0.7CaTiO$_{3}$-0.3NdAlO$_{3}$
  and Ba(Zn,Nb)O$_{3}$). Whilst the barium strontium titanate system
  (Ba$_{1-x}$Sr$_{x}$TiO$_{3}$) has been examined in detail
  experimentally
  \cite{bib:OxidesFerroelectricBaSrTiO3ForMicrowaveDevices,bib:FerroelectricsForMicrowaveApplications,bib:FerroelectricCeramicsBSTForMicrowaveApplications,bib:EffectOfSrTiOConcAndSinteringTempOnPropertiesOfBaTiO},
  it has not been manufactured and tested over the complete range from
  pure BaTiO$_{3}$ to pure SrTiO$_{3}$. The present paper describes
  the development of an ANN capable of predicting the relative
  permittivity of barium strontium titanate along with many other
  perovskite materials.

  Guo \emph{et al}. have previously investigated the use of ANNs for
  the prediction of the properties of dielectric ceramics such as
  BaTiO$_{3}$ \cite{bib:InvestigationOfBTANNMethod} etc. Their work
  concentrated on the effect of the addition of other compounds
  (lanthanum oxide, niobium oxide, samarium oxide, cobalt oxide and
  lithium carbonate) to pure barium titanate. Other work by Schweitzer
  \emph{et al}. \cite{bib:DevelopmentOfSPRForPredictionOfDCUsingNN}
  attempted prediction of dielectric data listed in the \emph{CRC
  Handbook} and the \emph{Handbook of Organic Chemistry}. This work
  used molecular information such as topological (bond type, number of
  occurrences of a structural fragment or functional group) and
  geometric (moment of inertia, molecular volume, surface area)
  descriptors in addition to the compositional information as the
  input variables. Additionally, there has been considerable work
  aimed at predicting the electrical properties of lead zirconium
  titanate (PZT) using ANN techniques
  \cite{bib:ModelingAndAnalysisOfPZTByNNs,bib:ApplicationOfANNsForDesignOfDielectricCeramics,bib:AnalysisOfPZTByANNs}.
  PZT is a piezoelectric ceramic material which finds increasing
  application in actuators and transducers.

  \subsection{Ion-diffusion materials for fuel cells}
  As noted above, ion-conducting ceramics are used in electrochemical
  devices such as oxygen separation membranes, solid oxide fuel cell
  (SOFC) cathodes and syngas reactors. Solid oxide fuel cells are of
  great interest as economical, clean and efficient power generation
  devices \cite{bib:OxygenIonSemiconductors}. Fuel cells have several
  advantages over conventional power generation techniques including
  their high-energy conversion efficiency and high power density while
  engendering extremely low pollution, in addition to the flexibility
  they confer in the use of hydrocarbon fuel
  \cite{bib:SynthesisOfLSCPowdersForSOFCCathodes}. Traditionally,
  large scale SOFCs have been based on yttria stabilised zirconia
  (YSZ) electrolytes and operate at high temperature (\degrees{1000}),
  placing considerable restrictions on the materials that can be
  used. Reduction of the operating temperature is essential for the
  future successful development of SOFCs, allowing increased
  reliability and the use of a wider range of materials.

  SOFC cathodes have stringent requirements. Ideally, the cathode
  material should be stable in an oxidising environment, have a high
  electrical conductivity, be thermally and chemically compatible with
  the other components of the cell and have sufficient porosity to
  allow gas transport to the oxidation site. Critically, the cathode
  material must allow diffusion of oxygen ions through the crystal
  lattice. The flexible perovskite structure of these materials allows
  doping, introducing defects into the lattice and facilitating the
  diffusion of ion species through the material. Materials currently
  under investigation include
  La$_{1-x}$Sr$_{x}$Mn$_{y}$Co$_{1-y}$O$_{3}$ (LSMC)
  \cite{bib:HighThroughputScreeningUsingSIMS},
  La$_{1-x}$Ca$_{x}$FeO$_{3-\delta}$ (LCF)
  \cite{bib:MicrostructureAndElectricalPropertiesOfLaFeCaO},
  La$_{2-x}$Sr$_{x}$NiO$_{4+\delta}$ (LSN)
  \cite{bib:OxygenDiffusionandSurfaceExchangeInLSN} and
  Ba$_{x}$Sr$_{1-x}$Co$_{1-y}$Fe$_{y}$O$_{3-\delta}$ (BSCF)
  \cite{bib:ThermalExpansionOfBSCF}. Much of the interest in these
  materials has stemmed from the fact that they form with oxygen
  deficiencies which provide a mechanism for fast oxygen ion transport
  through the defects in the crystal structure. Despite their ion
  transport properties, many possible SOFC cathode materials suffer
  from thermomechanical deficiencies such as cracking. Doping of Sr
  with other alkaline earth metals and replacing Mn, Co and Fe with
  other transition metals permits a wide range of possible materials
  allowing development of a material with optimal ion transport and
  thermomechanical properties
  \cite{bib:OxygenDiffusionandSurfaceExchangeInLSN}.

  There has been considerable investigation into the prediction of
  overall fuel cell performance using ANN techniques
  \cite{bib:ModellingFuelCellPerformanceUsingAI,bib:ANNSimulatorForSOFCPerformancePrediction,bib:EmpericalModellingOfPEMFCPerformanceUsingANNs,bib:HybridNNModelForPEMFCs},
  and some work on the modelling of diffusion properties has also been
  carried out
  \cite{bib:ModellingAndDatabaseSearchMOICsByCombinatorialMethods}.
  However, there has been little work on the ANN prediction of oxygen
  diffusion properties of the ceramic materials used as individual
  components of fuel cells although Ali \emph{et
  al} \cite{bib:StructurePropertyRelationshipSOFCCathodes} have
  recently investigated the structure-performance relationship of SOFC
  electrodes. Here, we present the results of our work on the
  development of ANNs for the prediction of the oxygen diffusion
  properties of ceramic materials. These networks may be subsequently
  included in the larger FOXD project \cite{bib:foxd}, allowing
  development of optimal SOFC cathode materials.

  \section{Artificial neural networks}
  Artificial neural networks can be used to develop functional
  approximations to data with almost limitless application
  \cite{bib:ANNsFundamentalsComputingDesignApplication,bib:NeuralComputingOfPropertiesOfRandomCompositeMaterials}. ANNs
  use existing data to learn the functional relationships between
  inputs and outputs. Unlike standard statistical regression
  techniques, ANNs make no prior assumption of the input-output
  relationship, a powerful advantage in their application to complex
  systems.

  ANNs are formed from individual processing units, or \emph{neurons},
  connected together in a network. The individual units are arranged
  into layers and the power of the neural computation comes from the
  interconnection between the layers of processing units. An
  individual unit consists of weighted inputs, a \emph{combination}
  function, an \emph{activation} function and one output. The outputs
  of one layer are connected to the inputs of the next layer to form
  the network topology. The performance of the network is determined
  by the form of the activation function, the training algorithm and
  by the network architecture. The selection of input data and
  architecture is a non-trivial process
  \cite{bib:SelectionOfInputsAndStructureOfFFNN,bib:ModelSelectionInNNs}
  and can have a large effect on the ultimate predictive abilities of
  the network. The individual units operate by evaluating the
  combination function which transforms the input and weight vectors
  into a scalar value. The output of the combination function is
  transformed through the activation function to give the neuron's
  ``state of activation''. The use of a non-linear activation function
  is responsible for the ability of the network to learn non-linear
  functions as a whole.

  For an ANN to be able to make predictions, it must be trained. The
  training process involves the application of a \emph{training
  dataset} to the network. The training algorithm is used to
  iteratively adjust the network's interconnection weights so that the
  error in prediction of the training dataset records is minimised and
  the network reaches a specified level of accuracy. The network can
  then be used to predict output values for new input data and is said
  to generalise well if such predictions are found to be accurate.

  \subsection{Multi-layer perceptron networks}
  In a multi-layer perceptron (MLP) network, the individual processing
  units are known as \emph{perceptrons} and they are usually arranged
  into three layers: input, hidden and output. Hecht-Nielsen proved
  that any continuous function can be approximated over a range of
  inputs by using a three layer feed forward neural network with
  back-propagation of errors \cite{bib:TheoryOfBackpropagationNN}.

  The number of neurons in the input and output layers is determined
  by the number of independent and dependent variables
  respectively. The number of hidden neurons is determined by the
  complexity of the problem and is often obtained by trial and error
  although evolutionary computing techniques such as genetic
  algorithms \cite{bib:PatternRecognitionUsingNeuralGeneticAlgorithm}
  have been used to determine optimal network architecture.

  We now describe a feed-forward neural network with back-propagation
  of errors. The operation of the network is as follows:

  \begin{enumerate}
  \item{Input some data $x_{i}$ to the input layer.}
  \item{Evaluate the combination function:

    \[
    c_{j} = \sum_{i}^{N} w_{ij}x_{i} + \theta
    \]

    \noindent which in this case is the \emph{dot product} of the
    input vector $x_{i}$ and the weights $w_{ij}$ where $j$ is
    the number of the hidden node being calculated, $\theta$ is the
    bias and $N$ is the length of the input vector.}
  \item{Calculate the value of the hidden node by applying a
    tanh-sigmoid activation function
    \[
    H_{j} = \frac{2}{(1+exp(-2c_{j}))}-1,
    \]
    where $j$ is the number of the hidden node and $y$ is the output
    of the combination function defined earlier.}
  \item{Calculate the network's output values $O_{k}$ at neuron $k$:
    \[
    O_{k} = g' \left( \sum_{l}^{P} w'_{lk}H_{l} + \theta' \right),
    \]
    where $k$ is the number of the output node being calculated,
    $\theta'$ is the bias, $w'_{lk}$ is the connection weight and $P$
    is the length of the hidden node vector (the number of hidden
    nodes); $g'$ is a linear activation function.}
  \item{Use the difference between $O_{k}$ and the data contained in
    the training set along with the derivative of the activation
    function to calculate the correction factor ($\delta_{k}$) to the
    weights connecting the hidden and output layer neurons: }
  \item{Use the correction factor to calculate the actual corrections
    to the weights connecting the hidden and output layer
    neurons
    \[
    w_{jk}^{new} = w_{jk}^{old} + \eta \delta_{k}H_{j}
    \]
    where $\eta$ is the \emph{learning rate} and controls the
    adjustments to the weights/biases.}
  \item{Calculate the correction factors for the weights connecting
    the input and hidden neurons, and insert these corrections}
  \item{Return to the first step and repeat the algorithm with the
    next entry in the training dataset.}
  \end{enumerate}

  The application of this algorithm to the complete training dataset
  is known as an \emph{epoch}. The network's performance is measured
  after each epoch has been completed and is determined by an error
  function. Two common error functions have been used, both based on
  the difference between the network's prediction
  and the expected values for the entire dataset. The first error
  function is the root mean square (RMS) of the prediction error:

  \begin{equation}
    \label{eqn:rms_error}
    \epsilon_{RMS} = \sqrt{\frac{\sum_{i=1}^{N}(y_{i} - t_{i})^2}{N}},
  \end{equation}

  \noindent where $y$ is the output predicted by the network, $t$ is
  the experimentally measured output and $N$ is the number of records
  in the dataset. The second error function is known as the root
  relative squared (RRS) error and is given by:

  \begin{equation}
    \label{eqn:rrs_error}
    \epsilon_{RRS} = \sqrt{\frac{\sum_{i=1}^{N}{(y_{i} -
	  t_{i})^2}}{\sum_{i=1}^{N}{(t_{i} - \overline{t})}}},
  \end{equation}

  \noindent where $\overline{t}$ is the mean of the experimentally
  measured outputs and the other symbols have been defined previously.

  The training process corresponds to an iterative decrease in the
  error function and continues until a predetermined value is reached,
  when training is halted. The trained network is tested through the
  application of previously unseen data to determine the
  performance. A network which performs well when working on new data
  is said to have good generalisation properties. As with statistical
  regression models, ANNs tend to perform much better when
  interpolating than extrapolating predictions. That is, whilst
  predictions are possible for any values of the input space, the most
  accurate and reliable results will be found when attempting
  predictions of materials which are similar to materials found in the
  training dataset.

  The selection of the error function value at which the training
  process is halted is not as simple as might first appear. The
  obvious choice is to select a low value, to obtain as high accuracy
  as possible. Unfortunately, this is found to lead to
  \emph{over-training}: the training dataset is ``memorised'' by the
  network and the generalisation to new data is poor. The effects of
  over-training occurring can be reduced by the use of another
  dataset, known as a \emph{validation dataset} which is used to
  monitor the training process. After each epoch of training, the
  network is used to predict the output values of the validation
  dataset and the error function (\ref{eqn:rms_error} or
  \ref{eqn:rrs_error}) of the validation dataset is calculated. When
  training starts, the error function of the validation dataset
  decreases in line with the error function of the training
  dataset. However, as the network begins to become over-trained, the
  error function of the validation dataset increases, and training is
  halted. It is the point where the error function of the validation
  dataset reaches a minimum that the network is expected to have the
  best generalisation performance. The use of the validation dataset
  to help prevent over-training is known as \emph{early stopping}
  \cite{bib:NeuralNetworksForPatternRecognition} of the training
  process.

  \subsection{Radial basis function networks}
  Radial basis function (RBF) neural networks \cite{bib:RBF} operate
  in a similar fashion to MLP networks. The key difference is that the
  combination function is the \emph{euclidean distance} between the
  input vector and the weight vector instead of the dot-product used
  in MLP networks. The most common form of basis function used is the
  Gaussian:

  \begin{equation}
  H_{j} = \exp{\left(-\frac{x^{2}}{2\sigma^{2}}\right)},
  \end{equation}

  \noindent where $x$ is the euclidean distance between the input
  vector and the centre of the Gaussian basis function and $\sigma$ is
  a parameter which determines the ``width'' of the function.

  RBF training algorithms operate in two stages. The first is
  unsupervised and uses only the input data of the training set. This
  stage involves the use of clustering algorithms such as K-means
  clustering \cite{bib:KMeansRBF} to determine suitable locations and
  width parameters for the basis functions. The second stage is
  identical to that used in MLP networks. Once the training algorithm
  has been used to locate the basis functions throughout parameter
  space and to calculate the second layer weights, the RBF network can
  be used to give predictions on new data. An important theoretical
  advantage of RBF over MLP networks is that the RBF training
  algorithm is the solution of a linear problem and can often be
  performed much faster than the complete non-linear optimisation
  required in the training of an MLP network.

  \subsection{Generalisation in artificial neural networks}
  The goal of ANN methods is to develop a network which is capable of
  accurately predicting output data values for records which are
  previously unseen by the network. An estimate of the generalisation
  performance of a network can be obtained by calculating the error
  function, eqn. (\ref{eqn:rms_error}) or (\ref{eqn:rrs_error}), of a
  dataset which is independent of that used for training. Such a
  dataset is known as the \emph{test dataset}. In order to utilise all
  of the data available, and to ensure that network performance is not
  simply due to coincidental dataset selection, cross-validation
  analysis is performed \cite{bib:MachineLearning}. In
  cross-validation, the data is divided into a number of subsets. All
  bar one of the datasets are employed for training/validation and the
  dataset withheld is used for testing. This process is repeated, each
  time withholding a different dataset and using the remainder for
  training/validation. In this way, all of the data is used for
  testing and the likelihood that the network performance is due to
  chance dataset selection is significantly reduced. Once complete,
  the mean of the error function from each repetition is
  calculated. This value is known as the \emph{generalisation error}
  and provides a measure of the overall performance of the network. To
  further increase confidence in the generalisation error, repeated
  cross-validation can be performed. In repeated cross-validation,
  cross-validation is performed several times, randomising the data in
  between each cross-validation. In this way, $n$ times $m$-fold
  cross-validation is performed, the network is trained $n$ x $m$
  times and we can be even more confident that the quoted
  generalisation error is accurate.

  \section{Ceramic materials datasets}
  Our dielectric dataset contains 700 records on the composition of
  dielectric resonator materials and their properties
  \cite{bib:DielectricDataset}. Many ceramic properties such as
  porosity, grain size, raw materials, processing parameters,
  measurement techniques and even the equipment used to manufacture
  them can all affect the dielectric properties. Since all material
  properties can be affected by such parameters the inclusion of such
  information may increase our ability to predict ceramic material
  properties.

  The majority of materials found in the dataset are Group II
  titanates, and Group II and transition metal oxides. Also included
  are some oxides of the lanthanides and actinides. The dataset
  contains relative permittivity values and Q-factors for 99\% of the
  records. Resonant frequency and temperature coefficient of resonant
  frequency data are also listed, but are only available for 58\% and
  83\% of the records respectively. The 700 records in the training
  dataset contain 53 different elements of which these materials may
  be comprised (Ag, Al, B, Ba, Bi, Ca, Cd, Ce, Co, Cr, Cu, Dy, Er, Eu,
  Fe, Ga, Gd, Ge, Hf, Ho, In, La, Li, M, Mg, Mn, Mo, Na, Nb, Nd, Ni,
  O, P, Pb, Pr, Sb, Sc, Si, Sm, Sn, Sr, T, Ta, Tb, Te, Ti, Tm, V, W,
  Y, Yb, Zn, Zr). It is the proportion of each of these elements found
  in the ceramic material which forms the input to the network.

  In addition to the full dataset described above, an ``optimised''
  dielectric dataset was obtained. This consisted of a subset of the
  data, selected by removing all glass material and all materials
  containing unusual dopants. The optimised dataset consists of 90
  records containing 37 different elements (Al, Ba, Bi, Ca, Ce, Co,
  Cu, Eu, F, Fe, Ga, Gd, Ge, Hf, La, Li, M, Mg, Mn, Na, Nb, Nd, Ni, O,
  Pb, Pr, Si, Sm, Sn, Sr, T, Ta, Ti, V, W, Zn, Zr). Again, the
  compositional information forms the input to the neural network and
  the dielectric properties the output.

  Our ion-diffusion dataset contains 1100 records of oxygen diffusing
  materials and their properties. The input data used for mining of
  the ion-diffusion data mainly consists of the compositional
  information of each material as in the dielectric dataset. The
  materials consist of Group II, transition metal, lanthanide and
  actinide oxides and contain 32 different elements (Al, Ba, Bi, Ca,
  Cd, Ce, Co, Cr, Cu, Dy, Fe, Ga, Gd, Ho, In, La, Mg, Mn, Nb, Nd, Ni,
  O, Pr, Sc, Si, Sm, Sr, Ti, V, Y, Yb, Zr). The proportion of these
  elements, along with the temperature at which the diffusion
  coefficient was measured from the network inputs. This dataset was
  collected from published sources. Unlike the records contained in
  the dielectric dataset, the ion-diffusion data contains many records
  which are measurements of the same material composition, performed
  at different temperatures. To differentiate between such
  measurements, the measurement temperature is included as an input
  variable of the ANN.

  \section{Neural network operation}
  Pre-processing of training data improves training stability and
  helps to prevent computational over- or underflow. All of the data
  is scaled so that the mean value is 0 and the standard deviation is
  1. In addition to the scaling algorithms, a technique called
  principal component analysis (PCA) is performed to
  remove any linear dependence of the input variables \cite{bib:PCA}.

  For the dielectric data, principal component analysis (PCA) was used
  to reduce the dimensionality of the dataset from the original 53
  elements to 16 by removing 2\% of the variation of the
  data. Similarly, for the optimised dielectric dataset, PCA reduced
  the dimensionality from 37 to 21. PCA of the ion-diffusion data
  allowed the dataset to be reduced from 33 elements to 16 by removing
  2\% of the variation of the data. The datasets used are randomly
  selected from the available data. The full set of data was split
  into three datasets: training, validation and test. As part of the
  cross-validation analysis, the data was divided into 10 equal size
  sub-datasets. One of the datasets is used for testing and the
  remainder is used for training and validation.

  The network contains three layers: input, hidden and output. The
  number of inputs and outputs were determined by the dimensions of
  the input data and the number of properties that we were aiming to
  predict. The number of hidden nodes was determined by trial and
  error and was chosen to be 15 for all three networks (dielectric,
  optimised dielectric and diffusion). When training was attempted
  with 10 hidden nodes, the network was not flexible enough to allow
  the network to learn the relationships, and generalisation was
  poor. The use of 20 hidden nodes gave a negligible performance
  increase. The computational requirements of the training process are
  low; on a 1.6 GHz single processor machine, the training of a 700
  record dataset was completed in 3600 epochs and took approximately 1
  minute. The ANNs were developed in Matlab \cite{bib:Matlab}, making
  extensive use of the Neural Network Toolbox
  \cite{bib:MatlabNNToolbox}.

  Initial attempts to train the neural network using the dielectric
  dataset resulted in poor generalisation. The dataset contains
  records with relative permittivities in the 0-1000 range. Especially
  poor results were obtained when attempting prediction of materials
  with permittivity greater than 100. Investigation revealed that the
  number of records with permittivity greater than 100 is far fewer
  than that in the range 0-100: 91\% of the records are in the 0-100
  range and the remaining 9\% in the range 100-1000. This resulted in
  the network being unable to accurately learn which material
  compositions produce relative permittivities greater than 100.

  Records associated with materials which exhibit relative
  permittivity greater than 100 were removed from the dataset. When
  network training was restarted, the performance of the network
  improved considerably, allowing accurate generalised predictions of
  the relative permittivity. However, as mentioned before, statistical
  techniques are more reliable when interpolating and so, whilst the
  predictive ability in the 0-100 range increased, extrapolation,
  predicting relative permittivity greater than 100, is likely to be
  relatively inaccurate.

  The diffusion coefficients of the data in the ion-diffusion dataset
  vary over a wide range ($\sim$ 4 orders of magnitude) and our
  initial training attempts resulted in extremely poor accuracy. The
  data was preprocessed by taking logarithms of the diffusion
  coefficients which reduced the absolute range of the output data and
  resulted in much improved ANN performance.

  \section{Results}
  The trained neural networks have been used to predict the properties
  of the materials in the test datasets which have then been compared
  to the experimental results. In addition, we have carried out
  cross-validation analysis of the data. The tables show data from 10
  repetitions of 10-fold cross-validation analysis. To measure the
  overall network performance, we have calculated both RMS and RRS
  error functions of the test datasets of the 10-fold cross-validation
  analysis and then calculated the mean of these error functions. The
  dataset was then re-randomised, and the 10-fold cross validation
  performed again. Once 10 randomisations were performed, the mean of
  the error functions of each cross validation was determined. The
  tables in this section show the results from each cross-validation
  and the overall mean and standard deviation of these results. The
  cross-validation ensures that the results are generalised throughout
  the entire dataset and the multiple randomisations ensure that the
  results are not due to coincidental randomisation. The overall
  ``mean of mean'' values of the error functions give a good
  indication of the generalisation error and provide the expected
  accuracy of predictions made using the neural networks.

  Finally, some analysis of the materials in each of the
  cross-validation datasets has been performed. We have attempted to
  provide a measure of the difference of the test dataset from the
  training/validation datasets. To calculate this figure, the mean
  composition of the test dataset and the combined training/validation
  datasets were calculated. We then calculated the RMS of the
  difference between the two mean values to show how the materials in
  the test dataset compare to the materials in the combined
  training/validation dataset. Test datasets which have a low mean
  composition difference from the training/validation datasets are
  more similar to the training/validation data and thus likely to
  perform better than test datasets with a large mean composition
  difference.

  \subsection{Prediction performance of the network trained using the dielectric dataset}
\begin{figure}
  \begin{center}
    \includegraphics[width=11cm]{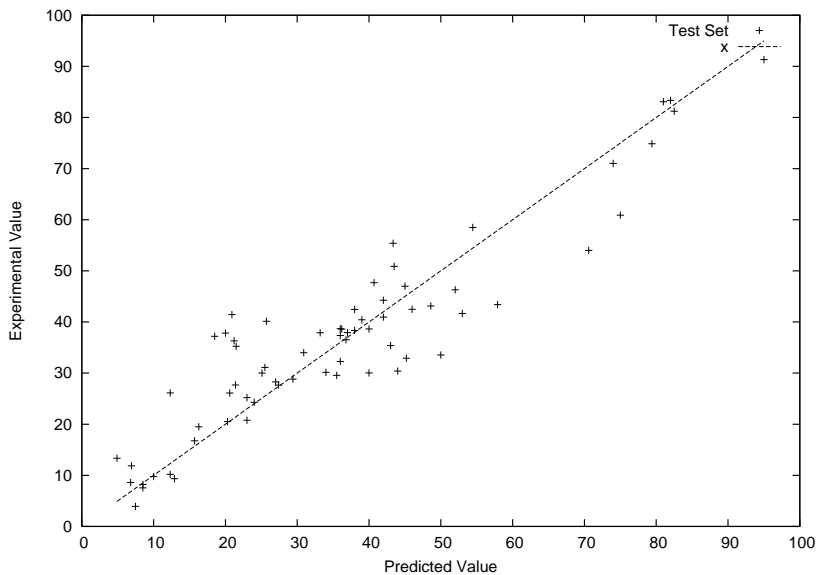}
    \caption{\textbf{The performance of the back-propagation MLP
	neural network used to predict the permittivity of the test
	dataset from the full dielectric dataset. This plot illustrates
	the performance of the second dataset combination in the
	cross-validation analysis (See Table
	\ref{tab:cross_validation_permittivity}). An ideal straight line
	with intercept 0 and slope 1 is also shown. The RRS error of the
	predictions is 0.61.}\label{fig:test_dataset}}
  \end{center}
\end{figure}

  The full dielectric dataset was divided into three sub-datasets
  (training, validation and test) and training was performed until
  halted by early stopping. The trained network was used to predict
  the (dimensionless) relative permittivity of the test dataset; the
  correlation between the experimentally observed permittivity and the
  predicted permittivity is shown in Figure \ref{fig:test_dataset}
  which demonstrates the accuracy of the predictions. The RMS error of
  the predicted data compared with the experimental data is
  0.61. Figure \ref{fig:test_dataset} is a plot of the second dataset
  combination from the cross-validation analysis.

  Statistical analysis of neural networks developed from the
  dielectric dataset was obtained by performing 10 repetitions of
  10-fold cross-validation analysis. Results of this analysis are
  provided in Table \ref{tab:cross_validation_permittivity} which
  shows the RMS and RRS error values, the parameters of a straight
  line fitted using least squares regression and the RMS of the mean
  compositional difference between the test dataset and the
  training/validation dataset. Also included are the the mean and
  standard deviation of these values. The values obtained are very
  similar as indicated by the standard deviation which confirms that
  each of the datasets contains a good representation of the whole
  dataset. This demonstrates that each sub-dataset is well randomised
  and the neural network performance is not simply due to the
  selection of the sub-datasets.

  Also shown is a repeated cross-validation analysis of the dielectric
  dataset with ionic radii data included (Table
  \ref{tab:cross_validation_permittivity_ionic_radius}). The ionic
  radius data was included by calculating the sum of the ionic radii
  of the elements in the corresponding material, in proportion to
  their fractional composition. The inclusion of ionic radius data
  leads to no change in the prediction performance of the network
  trained using the full dielectric dataset. The RRS error of the
  predictions remains at 0.6.

\begin{center}
\begin{table}
{\tiny
\begin{tabular}{ c c c c c c c c c c c c c }
\hline \multirow{2}{*}{\textbf{Quantity}} & \multicolumn{10}{c}{\textbf{Dataset randomisation}} & \multirow{2}{*}{\textbf{Mean}} & \multirow{2}{*}{\textbf{Std Dev.}}\\
 & 1 & 2 & 3 & 4 & 5 & 6 & 7 & 8 & 9 & 10 & & \\ \hline
Intercept & 1.05 & 1.62 & 0.27 & -0.25 & 2.33 & 0.75 & 0.22 & 1.44 & -0.88 & -0.02 & 0.65 & 0.97 \\ %\hline
Gradient & 0.98 & 0.96 & 0.98 & 1.01 & 0.96 & 0.97 & 1 & 0.97 & 1.03 & 0.99 & 0.99 & 0.02 \\ %\hline
Correlation & 0.63 & 0.63 & 0.68 & 0.65 & 0.64 & 0.62 & 0.64 & 0.65 & 0.65 & 0.63 & 0.64 & 0.02 \\ %\hline
RMS Error & 13.48 & 13.42 & 12.54 & 13.2 & 13.34 & 13.74 & 13.24 & 12.83 & 13.06 & 13.26 & 13.21 & 0.34 \\ %\hline
RMS mean material difference & 0.13 & 0.14 & 0.14 & 0.13 & 0.13 & 0.14 & 0.15 & 0.13 & 0.14 & 0.13 & 0.14 & 0.01 \\ %\hline
RRS Error & 0.62 & 0.62 & 0.57 & 0.6 & 0.61 & 0.62 & 0.6 & 0.58 & 0.59 & 0.6 & 0.6 & 0.02 \\ \hline
\end{tabular}
}
\textbf{\caption{The performance of the back-propagation MLP neural
          network used to predict the data within the test datasets
          taken from the dielectric dataset. Repeated cross-validation
          analysis was used to obtain these results and the mean and
          standard deviation are also
          given.\label{tab:cross_validation_permittivity}}}
\end{table}
\end{center}

\begin{center}
\begin{table}
{\tiny
\begin{tabular}{c c c c c c c c c c c c c}
\hline \multirow{2}{*}{\textbf{Quantity}} & \multicolumn{10}{c}{\textbf{Dataset randomisation}} & \multirow{2}{*}{\textbf{Mean}} & \multirow{2}{*}{\textbf{Std Dev.}}\\
 & 1 & 2 & 3 & 4 & 5 & 6 & 7 & 8 & 9 & 10 & & \\ \hline
Intercept & 0.73 & 0.39 & 0.96 & 0.75 & 1.57 & 1.36 & -0.65 & -0.02 & 2.21 & -1.29 & 0.6 & 1.05 \\ %\hline
Gradient & 0.99 & 0.98 & 0.99 & 0.97 & 0.95 & 0.96 & 1.01 & 1.00 & 0.96 & 1.01 & 0.98 & 0.02 \\ %\hline
Correlation & 0.65 & 0.67 & 0.65 & 0.63 & 0.62 & 0.62 & 0.67 & 0.64 & 0.67 & 0.68 & 0.65 & 0.02 \\ %\hline
RMS Error & 12.91 & 12.58 & 13.07 & 13.54 & 13.47 & 13.57 & 12.77 & 13.35 & 12.71 & 12.48 & 13.04 & 0.41 \\ %\hline
RMS mean material difference & 0.15 & 0.14 & 0.15 & 0.14 & 0.16 & 0.13 & 0.13 & 0.16 & 0.14 & 0.14 & 0.14 & 0.01 \\ %\hline
RRS Error & 0.59 & 0.58 & 0.6 & 0.62 & 0.63 & 0.61 & 0.58 & 0.60 & 0.58 & 0.57 & 0.6 & 0.02 \\ \hline
\end{tabular}
}
\textbf{\caption{The performance of the back-propagation MLP neural
	  network used to predict the data within the test datasets
	  taken from the dielectric dataset. The dataset includes
	  ionic radii as input variables. Repeated cross-validation
	  analysis was used to obtain these results and the mean and
	  standard deviation are also given. Comparison with the data
	  reported in Table \ref{tab:cross_validation_permittivity}
	  shows that inclusion of ionic radius has no effect on the
	  quality of
	  predictions. \label{tab:cross_validation_permittivity_ionic_radius}}}
\end{table}
\end{center}
  \subsection{Prediction performance of the network trained using the optimised dielectric dataset}
\begin{figure}
  \begin{center}
    \includegraphics[width=11cm]{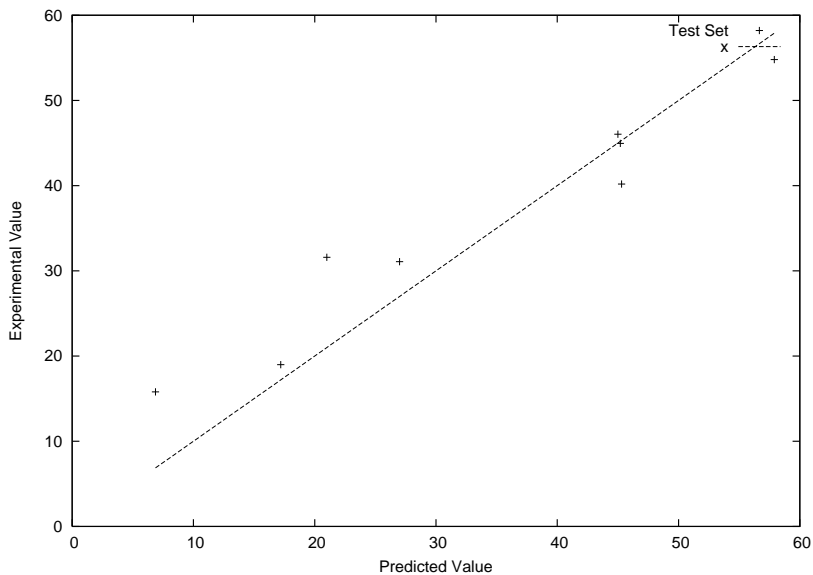}
    \textbf{\caption{The performance of the back-propagation MLP
	neural network used to predict the permittivity of the test
	dataset from the optimised dielectric dataset. This plot
	illustrates the performance of the first dataset in the
	cross-validation analysis (See Table
	\ref{tab:cross_validation_optimised}). An ideal straight line is
	shown as in the previous figure. The RRS error between
	experimental and predicted data is 0.63
	(dimensionless).}
      \label{fig:good_fit}}
  \end{center}
\end{figure}

  The optimised dielectric dataset was examined in a similar fashion
  to the full dielectric dataset. The dataset was divided into three,
  and training carried out using the early stopping technique to
  prevent over-training. Relative permittivity predictions of the test
  dataset were again obtained and the networks performance is
  summarised in Figure \ref{fig:good_fit}. This figure shows the
  accuracy of the neural network predictions compared to those
  obtained by experiment. The straight line shows the ideal
  correlation.

  As before, network training was performed using cross-validation
  analysis. The results of this are summarised in Table
  \ref{tab:cross_validation_optimised}. Again, since the statistical
  data are similar for each of the trained networks, the datasets each
  contain a good representation of the whole dataset and the result
  obtained in Figure \ref{fig:good_fit} is not simply due to the
  random selection of the datasets.

  Also shown is a repeated cross-validation analysis of the optimised
  dielectric dataset with ionic radius data included (Table
  \ref{tab:cross_validation_optimised_ionic}). As before, the ionic
  radius data was included by calculating the sum of the ionic radii
  of the elements in the material, in proportion to their fractional
  composition within the material. The inclusion of ionic radii data
  results in an increase in prediction performance as indicated by the
  RRS error decrease from 0.71 to 0.65.

  Whilst the ANN's predictions agree well with the experimental values
  in the dataset, it should be remembered that the network uses the
  experimental results as part of the training process and is
  therefore itself subject to the error in the experimental data. An
  ANN will never be able to provide predictions of properties which
  are more accurate than the error in the experimental measurements.
  Unfortunately, we do not have any error information for the
  dielectric data. Since the neural network uses experimental data in
  the training algorithm, the experimental error represents the
  intrinsic accuracy of the network. Overall, the network performs
  better when using the complete rather than the optimised
  dataset. When only compositional information is included, the RRS
  error of the cross-validated system is reduced from 0.71 to 0.60
  when the entire dataset is used. The standard deviation of the RRS
  error function obtained from the optimised dataset is larger than
  for the full dataset, possibly indicating that there is insufficient
  data for training the network when using the optimised dataset.

  As stated earlier, we expect the trained networks to perform well in
  interpolation, but less reliably in extrapolation. We can attempt to
  gauge the probability that the prediction of the properties of a
  material are accurate by measuring the ``distance'' of a material's
  composition from the hypothetical mean material. If a material is
  within, say, one standard deviation of the mean, the network is
  operating close to known parameter space and the predictions
  obtained are more likely to be accurate than materials which are
  ``further away'' in parameter (here composition) space.

\begin{center}
\begin{table}
{\tiny
\begin{tabular}{c c c c c c c c c c c c c}
\hline \multirow{2}{*}{\textbf{Quantity}} & \multicolumn{10}{c}{\textbf{Dataset randomisation}} & \multirow{2}{*}{\textbf{Mean}} & \multirow{2}{*}{\textbf{Std Dev.}}\\
 & 1 & 2 & 3 & 4 & 5 & 6 & 7 & 8 & 9 & 10 & & \\ \hline
Intercept & 2.24 & 7.03 & 0.94 & 3 & -3.18 & -4.24 & -0.41 & 1.16 & -10.35 & 2.27 & -0.15 & 4.78 \\ %\hline
Gradient & 0.94 & 0.85 & 0.96 & 0.91 & 1.05 & 1.14 & 0.97 & 0.88 & 1.26 & 1.02 & 1 & 0.13 \\ %\hline
Correlation & 0.64 & 0.44 & 0.62 & 0.6 & 0.61 & 0.67 & 0.6 & 0.51 & 0.63 & 0.6 & 0.59 & 0.07 \\ %\hline
RMS Error & 13.87 & 19.23 & 15.37 & 14.19 & 13.71 & 14.47 & 15.37 & 17.33 & 15.51 & 15.32 & 15.44 & 1.7 \\ %\hline
RMS mean material difference & 0.4 & 0.38 & 0.38 & 0.38 & 0.42 & 0.4 & 0.38 & 0.4 & 0.4 & 0.39 & 0.39 & 0.01 \\ %\hline
RRS Error & 0.63 & 0.89 & 0.71 & 0.69 & 0.63 & 0.62 & 0.71 & 0.76 & 0.69 & 0.72 & 0.71 & 0.08 \\ \hline
\end{tabular}
}
\textbf{\caption{The performance of the back-propagation MLP neural
      network used to predict the data within the test datasets taken
      from the optimised dielectric data. Repeated cross-validation
      analysis was used to obtain these results and the mean and
      standard deviation are also given.
	\label{tab:cross_validation_optimised}}}
\end{table}
\end{center}

\begin{center}
\begin{table}
{\tiny
\begin{tabular}{c c c c c c c c c c c c c}
\hline \multirow{2}{*}{\textbf{Quantity}} & \multicolumn{10}{c}{\textbf{Dataset randomisation}} & \multirow{2}{*}{\textbf{Mean}} & \multirow{2}{*}{\textbf{Std Dev.}}\\
 & 1 & 2 & 3 & 4 & 5 & 6 & 7 & 8 & 9 & 10 & & \\ \hline
Intercept & 2.01 & 11.17 & 1.67 & -6.28 & 0.14 & 5.26 & -13.31 & -9.05 & -2.14 & -3.17 & -1.37 & 7.1 \\ %\hline
Gradient & 0.96 & 0.75 & 0.89 & 1.09 & 0.99 & 0.91 & 1.31 & 1.2 & 1.02 & 1.07 & 1.02 & 0.16 \\ %\hline
Correlation & 0.64 & 0.56 & 0.57 & 0.69 & 0.71 & 0.57 & 0.57 & 0.64 & 0.73 & 0.73 & 0.64 & 0.07 \\ %\hline
RMS Error & 14.04 & 15.31 & 17.46 & 14.81 & 12.41 & 16.07 & 15.73 & 14.82 & 14.63 & 13.02 & 14.83 & 1.46 \\ %\hline
RMS mean material difference & 0.39 & 0.41 & 0.38 & 0.38 & 0.36 & 0.40 & 0.36 & 0.38 & 0.39 & 0.40 & 0.38 & 0.02 \\ %\hline
RRS Error & 0.61 & 0.70 & 0.74 & 0.63 & 0.53 & 0.75 & 0.68 & 0.62 & 0.65 & 0.55 & 0.65 & 0.07 \\ \hline
\end{tabular}
}
\textbf{\caption{The performance of the back-propagation MLP neural
	  network used to predict the data within the test datasets
	  taken from the optimised dielectric dataset. The dataset
	  includes ionic radius data as a input variable. Repeated
	  cross-validation analysis was used to obtain these results
	  and the mean and standard deviation are also
	  given.\label{tab:cross_validation_optimised_ionic}}}
\end{table}
\end{center}

  \subsection{Prediction performance of the network trained using the
    ion-diffusion dataset}

\begin{figure}
  \begin{center}
    \includegraphics[width=11cm]{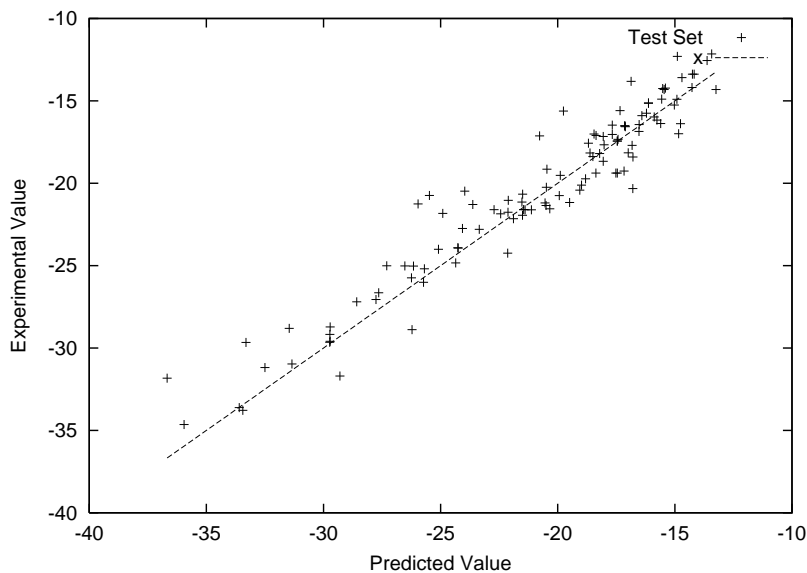}
    \textbf{\caption{The performance of the back-propagation MLP
	neural network used to predict the diffusion coefficient
	(cm$^{2}$s$^{-1}$) of the test dataset from the ion-diffusion
	dataset. The RMS error between experimental and predicted data
	is 0.34 (dimensionless, since the network is trained
	using the logarithm of the diffusion data).
	\label{fig:ic_performance}}}
  \end{center}
\end{figure}

  Analysis of the ion-diffusion dataset was performed using the same
  method as for the dielectric dataset. The dataset was randomised,
  divided into the three sub-datasets and training carried out until
  halted by the early stopping technique. The trained network was used
  to predict the logarithm of the diffusion coefficient
  (cm$^{2}$s$^{-1}$) of the records in the test dataset. The
  comparison between the predicted and experimental values is shown in
  Figure \ref{fig:ic_performance} and the RRS error of the predicted
  data compared to the experimental data is 2.12 (dimensionless since
  we are working with the logarithm of the diffusion coefficient).

  As for the dielectric dataset, it should be remembered that the
  network uses the experimental results as part of the training
  process and is subject to the error in the data. An ANN will never
  be able to provide predictions of properties which are more accurate
  than the error in the experimental measurements. Unfortunately, the
  ion-diffusion dataset only contains errors for about 3\% of the
  records. Due to the lack of error information, we are unable to
  perform comparisons between the ANN and experimental data and thus
  to determine whether or not the ANN predicts values within
  experimental error. As before, repeated cross-validation analysis
  was performed. The results of this are summarised in Table
  \ref{tab:cross_validation_diffusion}. The low standard deviation of
  the mean values shows that each of the datasets contains a good
  representation of the whole dataset and the result obtained in
  Figure \ref{fig:ic_performance} is not simply a coincidence of the
  randomisation and selection of the datasets. Again, interpolated
  predictions are more likely to be accurate than extrapolated results
  and we can use compositional distances from the mean composition to
  attempt to predict the expected accuracy of our predictions.

\begin{center}
\begin{table}
{\tiny
\begin{tabular}{c c c c c c c c c c c c c}
\hline \multirow{2}{*}{\textbf{Quantity}} & \multicolumn{10}{c}{\textbf{Dataset randomisation}} & \multirow{2}{*}{\textbf{Mean}} & \multirow{2}{*}{\textbf{Std Dev.}}\\
 & 1 & 2 & 3 & 4 & 5 & 6 & 7 & 8 & 9 & 10 & & \\ \hline
Intercept & -0.07 & -0.04 & -0.12 & 0.23 & -0.29 & 0.05 & -0.05 & 0.37 & 0.14 & 0.21 & 0.04 & 0.2 \\ %\hline
Gradient & 1 & 1 & 1 & 1.01 & 0.99 & 1.01 & 1 & 1.01 & 1.01 & 1.01 & 1 & 0.01 \\ %\hline
Correlation & 0.88 & 0.88 & 0.88 & 0.87 & 0.86 & 0.88 & 0.88 & 0.89 & 0.87 & 0.87 & 0.88 & 0.01 \\ %\hline
RMS Error & 2.12 & 2.07 & 2.1 & 2.13 & 2.26 & 2.08 & 2.1 & 2.04 & 2.14 & 2.15 & 2.12 & 0.06 \\ %\hline
RMS mean material difference & 0.11 & 0.11 & 0.11 & 0.1 & 0.11 & 0.11 & 0.11 & 0.12 & 0.12 & 0.11 & 0.11 & 0.01 \\ %\hline
RRS Error & 0.35 & 0.34 & 0.34 & 0.35 & 0.37 & 0.34 & 0.34 & 0.34 & 0.35 & 0.35 & 0.35 & 0.01 \\ \hline
\end{tabular}
}
\textbf{\caption{The performance of the back-propagation ANN on the
	  ion-diffusion dataset. Repeated cross-validation analysis was used to
	  obtain these results and the mean and standard deviation are
	  also given. \label{tab:cross_validation_diffusion}}}
\end{table}
\end{center}

\subsection{Radial basis function networks}
  In contrast to MLP networks detailed in the previous section,
  attempted training of radial basis function networks resulted in
  networks which generalised poorly. After making attempts to train
  networks using spherical RBFs, using the K-means clustering
  algorithm, we proceeded to modify the code to allow ellipsoidal
  basis functions which unfortunately resulted in no improvement. A
  possible reason for the failure of RBF networks to predict the
  materials properties in this study is that RBF networks perform
  poorly when there are input variables which have significant
  variance, but which are uncorrelated with the output variable
  \cite{bib:NeuralNetworksForPatternRecognition}. MLP networks learn
  to ignore the irrelevant inputs whilst RBF networks require a large
  number of hidden units to achieve accurate predictions.

  \section{Conclusions}

  Through application of artificial neural networks to pre-existing
  datasets culled from the literature, we have demonstrated that we
  can predict the permittivities and diffusion coefficients of ceramic
  materials simply from their composition and, in the case of the
  diffusion coefficient, experimental measurement temperature. A three
  layer perceptron network was trained using the back-propagation
  algorithm and cross-validation analysis of the data gave a mean root
  relative squared error of 0.6 for prediction of the dielectric
  constant of materials in the full dielectric dataset compared to
  0.71 for the smaller, optimised dataset. These results agree with
  previous work in the field of oilfield cements
  \cite{bib:UsingANNstoPredictCement} where neural network predictions
  were substantially enhanced when additional data records were
  included. The inclusion of ionic radius data results in no change to
  the prediction accuracy for the full dataset, although, a decrease
  in root relative squared error of 0.06 was found when the ionic
  radius data was included in the optimised dielectric dataset. The
  same network trained using the ion diffusion dataset was able to
  predict the logarithm of the oxygen diffusion coefficient with a RRS
  error of 0.35.

  Reliable Baconian methods for the prediction of the properties of
  ceramic materials are likely to become powerful tools for the
  scientific community whose accuracy will increase as more data is
  generated. The data produced by the FOXD project
  \cite{bib:InstrumentControlAndInformaticsSystem,bib:foxd} is
  beginning to accumulate and will be used to further develop these
  artificial neural networks. Through the use of evolutionary
  optimisation techniques such as the genetic algorithms of Holland
  \cite{bib:AdaptionInNaturalAndArtificialSystems}, we hope to be able
  to invert the neural networks described in this paper
  \cite{bib:UsingANNstoPredictCement}. This inversion provides the
  ability to search for materials with desirable properties which can
  then be synthesised using the London University Search
  Instrument. Other data mining tools including rule induction
  algorithms such as C4.5 \cite{bib:ProgramsForMachineLearning} can
  also be used to provide explicit, meaningful performance prediction
  rules from neural networks \cite{bib:ExtratingRulesFromNNs}.

  As part of the larger FOXD project, artificial neural networks akin
  to those developed here will form a vital link in the materials
  discovery cycle, leading to the possibility of steering automated
  searches in the compositional search space. In addition to producing
  data for further artificial neural network studies, ultimately we
  hope to use these techniques to discover and investigate new
  materials suitable for use in telecommunications, fuel cell and
  other areas.

  \section{Acknowledgements}
  We wish to thank Simon Clifford for fruitful discussions and Rob
  Pullar for dielectric property information. We would also like to
  express our thanks to Professor Julian Evans, Shoufeng Yang, Lifeng
  Chen and Yong Zhang at Queen Mary College, University of London.

  This research is supported by the EPSRC-funded project ``Discovery
  of New Functional Oxides by Combinatorial Methods'' (GR/S85269/01)
  \cite{bib:foxd}. Copies of the software and literature datasets
  described herein may be obtained upon application to the authors.

  \bibliography{bibliography}
  \bibliographystyle{elsart-num}

\end{document}